\documentclass[twocolumn,showpacs,preprintnumbers,amsmath,amssymb,showkeys]{revtex4}

\usepackage{graphicx}
\usepackage{dcolumn}
\usepackage{bm}

\begin{document}

\preprint{APS/123-QED}

\title{Toroidal metamaterial} 

\author{K. Marinov$^{1,2,}$}%
 \email{k.b.marinov@dl.ac.uk}

\author{A. D. Boardman$^{1}$}
\affiliation{%
$^{1}$Photonics and Nonlinear Science Group, Joule Laboratory, Department of Physics,
University of Salford, Salford M5 4WT, UK
}%

\author{V. A. Fedotov$^{2}$}
\author{N. Zheludev$^{2,}$}
 \email{niz@orc.soton.ac.uk}

\affiliation{%
$^{2}$EPSRC Nanophotonics Portfolio Centre, School of Physics and
Astronomy, University of Southampton, Highfield, Southampton, SO17
1BJ, UK
}%

\date{\today}

\begin{abstract}
It is shown that a new type of metamaterial, a 3D-array of toroidal solenoids, displays a significant toroidal response that can be readily measured. This is in sharp contrast to materials that exist in nature, where the toroidal component is weak and hardly measurable. The existence of an optimal configuration, maximizing the interaction with an external electromagnetic field, is demonstrated. In addition, it is found that a characteristic feature of the magnetic toroidal response is its strong dependence on the background dielectric permittivity of the host material, which suggests possible applications. Negative refraction and backward waves exist in a composite toroidal metamaterial, consisting of an array of wires and an array of toroidal solenoids. 
\end{abstract}

\maketitle


Toroidal moments were first considered by Ya. Zel'dovich in 1957 \cite{1}. They are fundamental electromagnetic excitations that cannot be represented in terms of the standard multipole expansion \cite{2}. The properties of materials that possess toroidal moments and the classification of their interactions with external electromagnetic fields have become a subject of growing interest \cite{2, 3, 4, 5, 6, 7, 8, 9, 10, 11, 12, 13, 13a, 13b}. In particular the unusual properties of non-radiating configurations, based on toroidal and supertoroidal currents, have been discussed \cite{4, 5, 10}. Non-reciprocal refraction associated with an effective Lorentz force acting on photons propagating in a toroidal domain wall material has recently been predicted \cite {7}. Whereas the importance of toroidal moments in particle physics has been established some time ago (see \cite{12} and the references therein), there is no known observation of toroidal response in the classical electromagnetism \cite{9}. This is because the effects associated with toroidal response in materials  \textit{that exist in nature} are weak \cite{7,8,9}. At the same time modern technology allows arrays of toroidal moments with sub-milimeter particle-size to be manifactured and characterized \cite{13}. Therefore, a metamaterial specifically designed to maximize the \textit{toroidal component} of an interaction with an external electromagnetic field may provide an outcome that can open up new possibilities for observation of toroidal response and its applications. 
\begin{figure}
\includegraphics{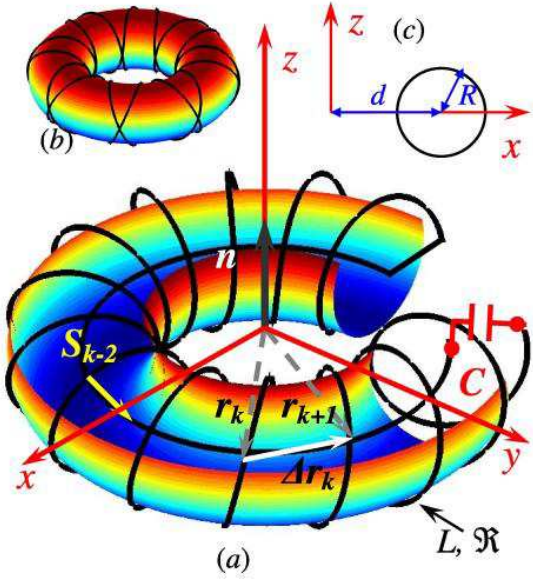}
\caption{\label{fig:fig1} (\textit{a}) Toroidal moment - a toroid is wound continuously with wire and an extra loop of radius $R_{1}=d\sqrt{1+0.5(R/d)^{2}}$, compensating the magnetic dipole moment of the toroidal section is introduced in the equatorial plane. $L$ is the inductance of the solenoid and $\Re$ is its loss resistance. $C$ is an externally introduced capacitor.  The origin of the co-ordinate system is the geometric center of the toroid. $\bm{S_{k}}\approx S(\Delta \bm{r_{k}}/|\Delta \bm{r_{k}}|)$, where $S=\pi R^{2}$ and $|\bm{r_{k}}|=d$. (\textit{b}) A double-wound toroidal solenoid has a zero net magnetic dipole moment.  (\textit{c})  $d$ and $R$ are the larger and the smaller radii of the toroidal surface $(\sqrt{x^2+y^2}-d)^{2}+z^{2}=R^{2}$, respectively.}  
\end{figure}

The aim of the present paper is to provide an estimate for the magnitude of the response of an array of toroidal moments to an external electromagnetic field. The model is valid in a quasistatic regime (the size of each metaparticle is much smaller than the wavelength) and with the assumption that a \textit{uniform} current is induced in each metaparticle. It is shown that a toroidal  metamaterial - an array of toroidal solenoids - shows a significant electromagnetic response and has intriguing electromagnetic properties.  A characteristic feature is the strong dependence of the response bandwidth on the dielectric properties of the host material, which suggests possible applications.  Backward waves and negative refraction exist in a composite toroidal material consisting of an array of wires and an array of toroidal metaparticles.  

It should be emphasized that the term "anapole" (often used in nuclear physics problems) and the term "toroidal moment"  refer to physically different objects (see the relevant discussion in \cite{11}). Indeed, an anapole does not radiate electromagnetic energy and does not interact with external electromagnetic fields. In contrast, toroidal moments interact with \textit{time-dependent} electromagnetic fields and radiate (scatter) electromagnetic energy.  Hence, an array of toroidal moments can act as an electromagnetic medium in the same way in which an array of split-ring resonators does.  

The discussion begins by considering a metaparticle in the form a toroidal solenoid as shown in Fig. \ref{fig:fig1}. A conducting wire is wrapped around the toroid so that there are $N$ turns between the entrance and the exit points for the uniform current flowing through the wire. The two ends of the wire are connected through a capacitor $C$ and through an additional loop in the equatorial plane of the structure to compensate the $z$-component of the magnetic dipole moment associated with the azimuthal component of the current flowing in the toroidal section of the winding. An alternative way to eliminate the latter is to use double winding, as shown in Fig. \ref{fig:fig1}(\textit{b}) \cite{13b}. It will be shown that the magnetic \textit{quadrupole} moments of both the structures shown in Fig. \ref{fig:fig1} are equal to zero. 

Indeed, the toroidal dipole moment of a current density distribution $\bm{j}(\bm{r},t)$ is a \textit{vector} given by \cite{2, 12, 13a}
\begin{eqnarray}
    \bm{T}=\frac{1}{10}\int\left[\left(\bm{j}.\bm{r}\right)\bm{r}-2r^{2}\bm{j}\right]dV.
    \label{eq:zero}
\end{eqnarray}
On the other hand, the magnetic quadrupole moment  of the same system is the  \textit{second-rank tensor} \cite{13a, 13c}
\begin{eqnarray}
    \hat{\bm{Q}}=\frac{1}{3}\int\left[\left(\bm{r}\times\bm{j}\right)\bm{r}+\bm{r}\left(\bm{r}\times\bm{j}\right)\right]dV.
    \label{eq:zero_a}
\end{eqnarray}
A toroidal coil, assumed to be made of infinitely thin wire can be generated using the following equations
\begin{subequations}       
\label{eq:zero_b}    
		\begin{equation}
			x=(d-R\cos N\varphi)\cos\varphi, 
  	\end{equation}
  	\begin{equation}
  		y=(d-R\cos N\varphi)\sin\varphi, 
		\end{equation}
		\begin{equation}
			z=R\sin N\varphi. 
		\end{equation}
\end{subequations}
If $\bm{r}=(x, y, z)$ is the radius-vector of an arbitrary point along the winding then $\varphi$ is the angle between the $x$-axis and the projection of $\bm{r}$ onto the $xy$-plane. The assumption of a uniform current $I$ flowing along the wire leads to the current element $\bm{j}dV$  being defined as $\bm{j}dV=I\bm{dr}$. Hence, given that $\bm{dr}=\frac{\bm{dr}}{d\varphi}d\varphi$, equations (\ref{eq:zero}) and (\ref{eq:zero_a}) become
\begin{eqnarray}
    \bm{T}=\frac{I}{10}\int^{2\pi}_{0}\left[\left(\frac{d\bm{r}}{d\varphi}.\bm{r}\right)\bm{r}-2r^{2}\frac{d\bm{r}}{d\varphi}\right]d\varphi
    \label{eq:zero_c}
\end{eqnarray}
and
\begin{eqnarray}    \hat{\bm{Q}}=\frac{I}{3}\int^{2\pi}_{0}\left[\left(\bm{r}\times\frac{d\bm{r}}{d\varphi}\right)\bm{r}+\bm{r}\left(\bm{r}\times\frac{d\bm{r}}{d\varphi}\right)\right]d\varphi,
    \label{eq:zero_d}
\end{eqnarray}
respectively. Evaluating $\frac{\bm{dr}}{d\varphi}$ from (\ref{eq:zero_b}), performing the integrations in (\ref{eq:zero_c}) and (\ref{eq:zero_d}) and assuming $N>1$ yields,  
\begin{eqnarray}
		\bm{T}=\frac{\pi N I d R^{2}}{2}\bm{n}    
    \label{eq:zero_e}
\end{eqnarray}
and
\begin{eqnarray}
		\hat{\bm{Q}}=0,  
    \label{eq:zero_f}
\end{eqnarray}
where $\bm{n}=(0, 0, 1)$ \cite{13b}. As (\ref{eq:zero_e}), (\ref{eq:zero_f}) clearly show, the quadrupole moment of a toroidal solenoid is zero whereas its toroidal moment is non-zero. Although, as (\ref{eq:zero}) and (\ref{eq:zero_a}) suggest, $\bm{T}$ and $\hat{\bm{Q}}$ are measured in the same units $[Am^{3}]$, they are physically different and this is reflected in their properties. For example, a DC toroidal moment $(I=const(t))$ does not generate magnetic field outside its own volume \cite{13b}, whereas a DC quadrupole moment generates non-zero magnetic field everywhere. In addition, the radiation pattern of an AC toroidal moment is the same as that of an electric dipole \cite{10, 13b}, which is not the case for a magnetic quadrupole moment.
The magnetic dipole moment of the toroidal coil 
\begin{eqnarray}
		\bm{m}=\frac{1}{2}\int\left(\bm{r}\times\bm{j}\right)dV=\frac{I}{2}\int^{2\pi}_{0}\left(\bm{r}\times\frac{d\bm{r}}{d\varphi}\right)d\varphi
    \label{eq:zero_g}
\end{eqnarray}
is
\begin{eqnarray}
		\bm{m}=\pi d^{2}I\left[1+\frac{1}{2}\left(\frac{R}{d}\right)^{2}\right]\bm{n} 
    \label{eq:zero_h}
\end{eqnarray}
and, hence, this magnetic moment can be compensated by a loop of radius $R_{1}=d\sqrt{1+0.5(R/d)^{2}}$ in the equatorial plane of the torus, carrying current uniform current $-I$, as shown in Fig. \ref{fig:fig1}(a). Note, that the magnetic quadrupole moment of this loop is zero and, hence, the magnetic quadrupole moment of the coil shown in Fig. \ref{fig:fig1}(a) is identically zero. 

A toroidal coil, wound in a direction opposite to that of the coil defined by the system (\ref{eq:zero_b}) relies instead upon the equations 
\begin{subequations}       
\label{eq:zero_i}    
		\begin{equation}
			x'=(d-R\cos N\varphi)\sin\varphi, 
  	\end{equation}
  	\begin{equation}
  		y'=(d-R\cos N\varphi)\cos\varphi, 
		\end{equation}
		\begin{equation}
			z'=R\sin N\varphi,
		\end{equation}
\end{subequations}
which are obtained from (\ref{eq:zero_b}) by formally exchanging $x$ and $y$ in the latter. The toroidal, magnetic dipole and magnetic quadrupole moments are now  $\bm{T'}=\bm{T}$, $\bm{m'}=-\bm{m}$ and  $\hat{\bm{Q}}'=0$, so that it can be concluded that a double-wound toroidal solenoid has a zero net magnetic dipole and magnetic quadrupole moments, whereas the toroidal moment of the structure is twice that of a single-wound torus (Fig. \ref{fig:fig1}b). The possibility of eliminating the magnetic dipole moment of a toroidal coil by either adding a loop in the equatorial plane of the torus, or using a double winding solenoid has been discussed earlier \cite{13b}. 

If either of the metaparticles shown in Fig. \ref{fig:fig1} is placed in an external \textit{inhomogeneous} magnetic field $\bm{B(r)}$ and the characteristic length scale of variation of the field is much larger than the size of the metaparticle, then the leading-order response to the external field comes from the toroidal moment. Indeed, since the current $I$ is uniform there is no charge accumulation and the electric multipole moments are zero. The magnetic dipole moment of the system can be elliminated as shown in Fig. \ref{fig:fig1} and the magnetic quadrupole moment is zero, as it has been demonstrated. Therefore, under these conditions it is toroidal moment $\bm{T}$ that generates the leading-order electromagnetic response of the system \cite{2, 12, 13a}.
The net magnetic flux $\varPhi$ in the solenoid can be written in the approximate form
\begin{eqnarray}
    \varPhi=\sum^{N}_{k=1}\bm{B_{k}}.\bm{S_{k}},
    \label{eq:one}
\end{eqnarray}
where $\bm{B_{k}}=\bm{B(r_{k})}$ is the flux density in the center of a given loop. Since $\bm{S_{k}}$ is a vector pointing along the normal to this loop, then $\bm{S_{k}}$ can be approximated to $\bm{S_{k}}\approx S(\Delta \bm{r_{k}}/|\Delta \bm{r_{k}}|)$, where $S=\pi R^{2}$, provided that $N$ is sufficiently large. Now $|\Delta \bm{r_{k}}|\approx 2 \pi d/N$, hence the flux (1) is 
\begin{eqnarray}
    \varPhi \approx \frac{SN}{2\pi d} \sum^{N}_{k=1}\bm{B_{k}}.\Delta \bm{r_{k}}\approx \frac{SN}{2\pi d}\oint_{\partial\varSigma}\bm{B}.\bm{dr}.
    \label{eq:two}
\end{eqnarray}
In (\ref{eq:two}) the discreet distribution of loops has been approximated with a continuous one, which is valid for a sufficiently large $N$ and $\partial\varSigma$ is a circle of radius $d$ centered at the origin. Using Stokes' theorem (\ref{eq:two}) can be rewritten as 
\begin{eqnarray}
    \varPhi=\frac{SNd}{2}(\bm{n}.[curl\;\bm{B}]_{\bm{r}=0}).
    \label{eq:three}
\end{eqnarray}
where once again the assumption for a weak magnetic field variation has been used. 
In (\ref{eq:three}) $[curl\;\bm{B}]_{\bm{r}=0}$ stands for $curl\;\bm{B}$ evaluated at $\bm{r}=0$.

The net energy $W$ of a toroidal metaparticle in an external magnetic field is the sum of the contributions of each loop, hence
 \begin{eqnarray}
    W=-I\sum^{N}_{k=1}\bm{B_{k}}.\bm{S_{k}},
    \label{eq:four}
\end{eqnarray}
where $I$ is the strength of the current flowing in the solenoid. From (\ref{eq:one}), (\ref{eq:three}) and (\ref{eq:four}) $W$ takes the standard form $W=-\bm{T}.curl\bm{B}$ \cite{2,3,4,5,11}, where  $\bm{T}$ is given by (\ref{eq:zero_e}).
\begin{figure}
\includegraphics{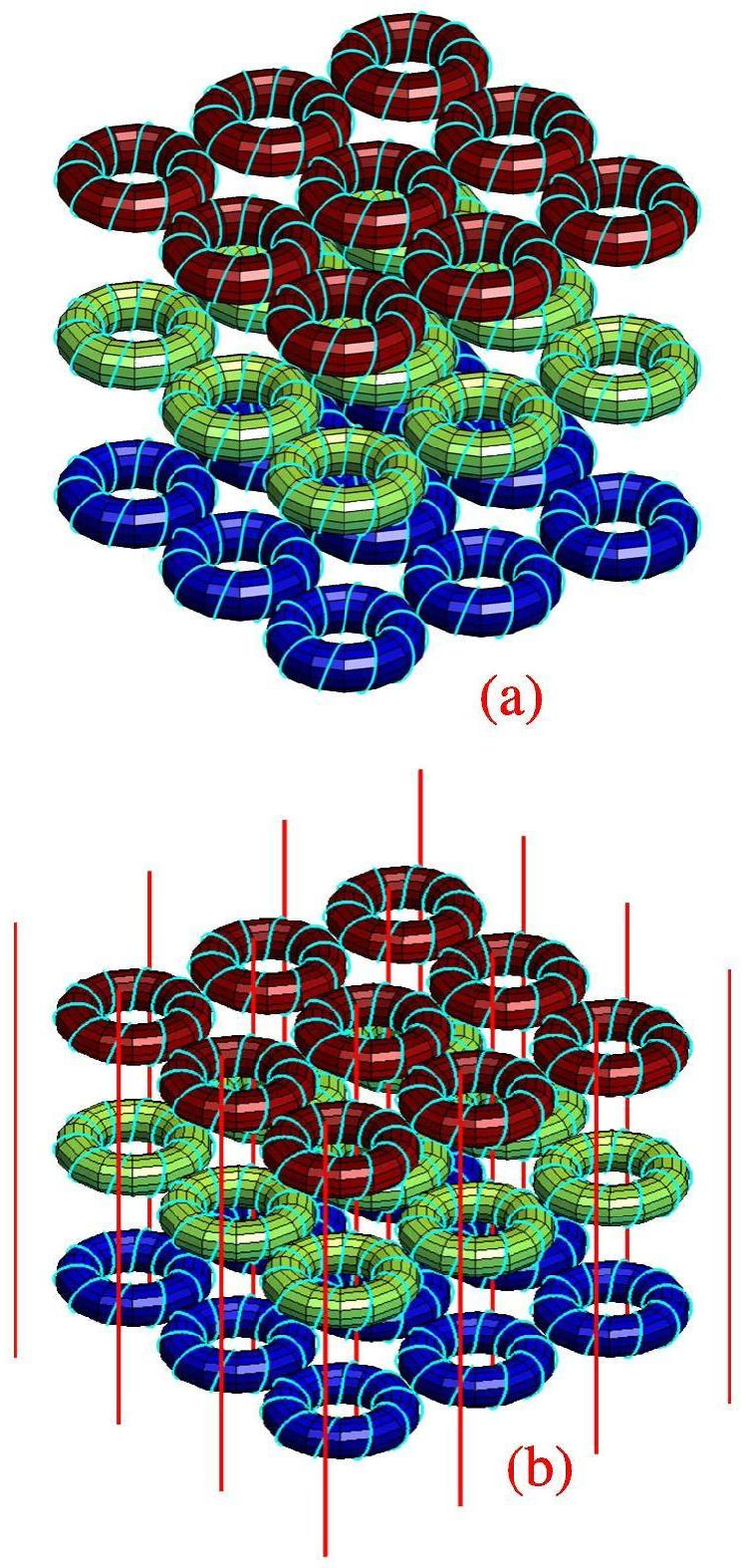}
\caption{\label{fig:fig2} (a) Toroidal metamaterial - array of toroidal solenoids embedded in an isotropic dielectric material with dielectric permittivity $\varepsilon$. (b) Composite toroidal metamaterial - array of toroidal solenoids and an array of long wires, parallel to the axis of the toroids.}
\end{figure}
Suppose that the external electromagnetic field acting on the metaparticle is that of a monochromatic plane wave $(\bm{E, H})\propto \exp(-i \omega t+i\bm{k.r})$, where $\bm{E}$ is the electric field, $\bm{H}$ is the magnetic field, $\omega$ is the angular frequency and $\bm{k}$ is the wavevector.  Kirchhoff's law implies that
\begin{eqnarray}
    L\frac{dI}{dt}+I\Re+\frac{Q}{C}=-\frac{d\varPhi'}{dt}.
    \label{eq:six}
\end{eqnarray}
In (\ref{eq:six}) $L$ and $\Re$ are the inductance and the resistance of the solenoid, respectively, $Q$ is the electric charge accumulated by the capacitor and $\varPhi'$ is the flux (\ref{eq:three}), generated by the local (microscopic) field $\bm{H'}$. The complex amplitude of the current $\widetilde{I}$ is, therefore, 
\begin{eqnarray}
    \widetilde{I}=\frac{i\omega \tau \mu_{0}}{i((\omega C)^{-1}-\omega L)+\Re}(curl\;\bm{H'}).\bm{n},
    \label{eq:seven}
\end{eqnarray}
where $\tau=(SNd)/2$ and $\mu_{0}$ is the vacuum permeability.
\begin{figure}
\includegraphics{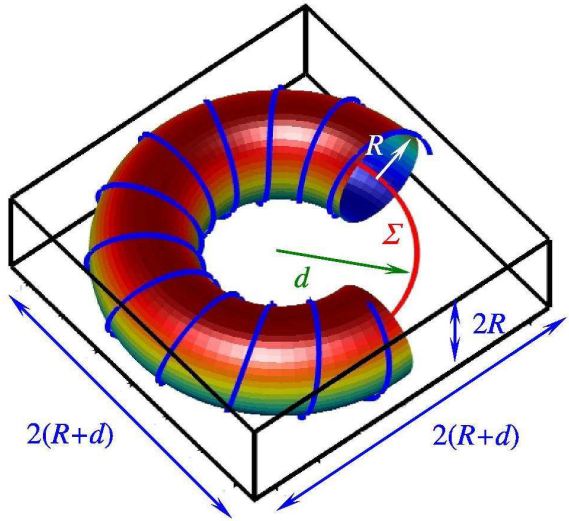}
\caption{\label{fig:fig3} The unit cell of a toroidal metamaterial. The minimum volume required for each toroid is $8R(d+R)^{2}$.}
\end{figure}
\begin{figure}
\includegraphics{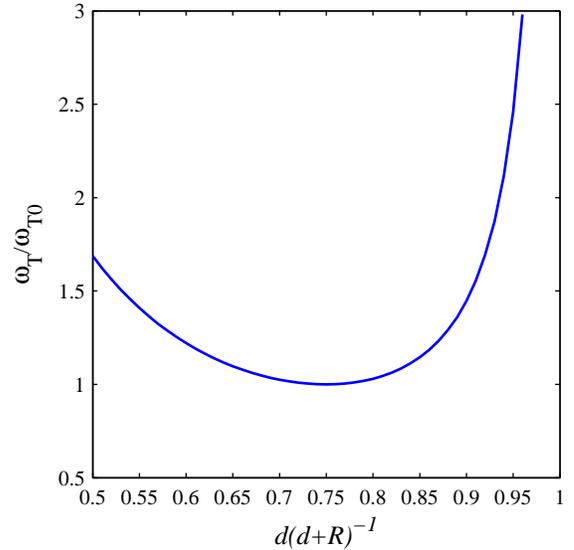}
\caption{\label{fig:fig4} Dependence of the toroidal frequency $\omega_{T}$ on the geometry of the unit cell. The transverse size $R+d$ is fixed and the ratio $R/d$ is the free parameter.}
\end{figure}
Consider now a toroidal metamaterial, a 3D array of toroidal metaparticles (Fig. \ref{fig:fig2}) embedded in a background medium. Two different types of host media are considered: ordinary isotropic dielectric media and arrays of long wires. The former situation is referred to as a "toroidal metamaterial" and the latter - a "composite toroidal metamaterial". It should be emphasized that if the wires do not cross the toroids, which is the situation depicted in Fig. \ref{fig:fig2}(b), no interaction between the neighboring wires and toroids is possible. This outcome flows directly from the expression for the interaction energy of a toroidal moment with an external field. Indeed, since $W=-\bm{T}.curl\bm{B}=-\mu_{0}\bm{T}.\bm{j}$, where $\bm{j}$ is the source of the field $\bm{B}$, it is clear, that $W$ is zero unless the locations of the toroidal moment and the field source coincide.   
The external field induces in each solenoid a \textit{time-dependent} toroidal moment $\bm{T}$ that can be obtained from (\ref{eq:zero_e}) and (\ref{eq:seven}). In a similar way to the description of an array of magnetic dipoles in terms of its effective magnetization, an array of toroids can be characterized by an effective toroidization vector $\bm{\varTheta}=\kappa \bm{T}$  where  $\kappa$ is the number of toroids per unit volume.  The macroscopic magnetization $\bm{M}$ of the material is then \cite{2,4,5,11}
\begin{eqnarray}
    \bm{M}=\nabla\times\bm{\varTheta}.
    \label{eq:eight}
\end{eqnarray}
Using (\ref{eq:zero_e}), (\ref{eq:seven}) and the Lorenz-Lorentz formula $\bm{H'}=\bm{H}+\bm{M}/3$ allows (\ref{eq:eight}) to be written as
\begin{eqnarray}
    \bm{M}=\frac{\omega^{2}c^{2}(\bm{k}\times\bm{H}).\bm{n} }{\omega^{2}_{T}\left[\omega^{2}\left(1+\frac{k_{\bot}^{2}c^{2}}{3\omega_{T}^{2}}\right)-\omega^{2}_{0}+i\gamma \omega\right]}(\bm{k}\times\bm{n}),
    \label{eq:nine}
\end{eqnarray}
where the effective "toroidal" frequency $\omega^{2}_{T}=Lc^{2}(\mu_{0}\tau^{2}\kappa)^{-1}$ has been introduced, $\omega_{0}=(LC)^{-1/2}$ is the resonant frequency, $\gamma=\Re/L$ is the damping factor,  $c$ is the speed of light in vacuum and $k_{\bot}^{2}=k^{2}-(\bm{k}.\bm{n})^{2}$. The toroidal frequency  plays a role similar to that of the effective "plasma" frequency $\omega_{p}$  in metamaterials involving arrays of wires \cite{14}.  	
Equation (\ref{eq:nine}) clearly shows that the metaparticles are not responding like magnetic dipoles. Indeed, the magnetic flux density vector in the constitutive relationships is not proportional to the magnetic field \textit{locally}, as is the case with metamaterials involving split-ring resonators (SRRs) \cite{14,15,16}. 
As (\ref{eq:nine}) shows, in order to maximize the electromagnetic response of the material one needs to minimize the toroidal frequency $\omega_{T}$. Given $R$ and $d$ the maximum value of the number density is $\kappa=(8R(d+R)^2)^{-1}$ (see Fig. \ref{fig:fig3}) and this corresponds to each toroid in Fig. \ref{fig:fig2} being in contact with its nearest neighbors. The toroidal frequency then is given by   
\begin{eqnarray}
    \omega_{T}^{2}=\frac{16c^2(d+R)^2}{\pi^{2}Rd^{3}}
    \label{eq:ten}
\end{eqnarray}
and it does not depend on the number of windings $N$. Note, that in deriving (\ref{eq:ten}) $L=\mu_{0}N^{2}R^{2}/(2d)$ has been used \cite{13}. If the transverse size of the unit cell $2(R+d)$ is fixed (e.g. by the manifacturing process), the ratio $R/d$ remains a free parameter and it is easy to show that $\omega_{T}$ reaches a minimum $\omega_{T0}=16c/(\pi d \sqrt{3}) $ at $d=3R$ and this value represents a \textit{fundamental} upper limit for the strength of the toroidal metamaterial response in the quasistatic regime, given the transverse size of the unit cell $2(R+d)$. As Fig. \ref{fig:fig4} shows the function $\omega_{T}/\omega_{T0}$ depends weakly on $d/R$ in a relatively broad range of $d/R$ values. However, structures with $R<<d$ are disadvatageous since $\omega_{T}\rightarrow\infty$ as $R\rightarrow0$. It should be noted that the inductance of the solenoid is given by $L=\mu_{0}N^{2}R^{2}/(2d)$ provided that two conditions are met: (i) the torus is closely wound, i.e. $N$ is sufficiently large and (ii) $R<<d$. The first condition has already been imposed in deriving (\ref{eq:two}) and, hence, its use in (\ref{eq:ten}) does not impose any further restrictions. Excellent agreement with experimental measurements has been obtained at $N=15$ \cite{13}. The inductance of a closely wound toroidal solenoid with arbitrary $R$ and $d$ is given by $L'=\mu_{0}N^{2}\left(d^{2}-\sqrt{d^{2}-R^{2}}\right)$ \cite{16a}. Note, however, that at $d=3R$, which is the physically interesting parameter range, the relative difference between $L$ and $L'$ is $|L'-L|/L<0.03$. Thus, the use of $L'$ instead of $L$ in  (\ref{eq:ten}) results in shifting the minimum of $\omega_{T}$ from $d=3R$ to $d\approx3.2R$.

With $d=$5 mm and $R=$1.7 mm the value of the toroidal frequency is $\omega_{T0}/2\pi=$28.3 GHz. In contrast, the micro-structures reported in \cite{13} have $\omega_{T0}/2\pi \approx $380 GHz. As (\ref{eq:ten}) suggests, the validity of the quasistatic approximation requires  $\omega<<\omega_{T}$. This is because $\omega_{T0}d/c \approx 3$ and for this reason in what follows $\omega\leq\omega_{T}/10$ is used. In addition, numerical simulations show that the leading order toroidal moment $\bm{T}$, given by  Eq.(\ref{eq:zero_e}), describes the electromagnetic properties of toroidal solenoids with $kd\approx0.2$ with a high degree of accuracy \cite{10}.
\begin{figure}
\includegraphics{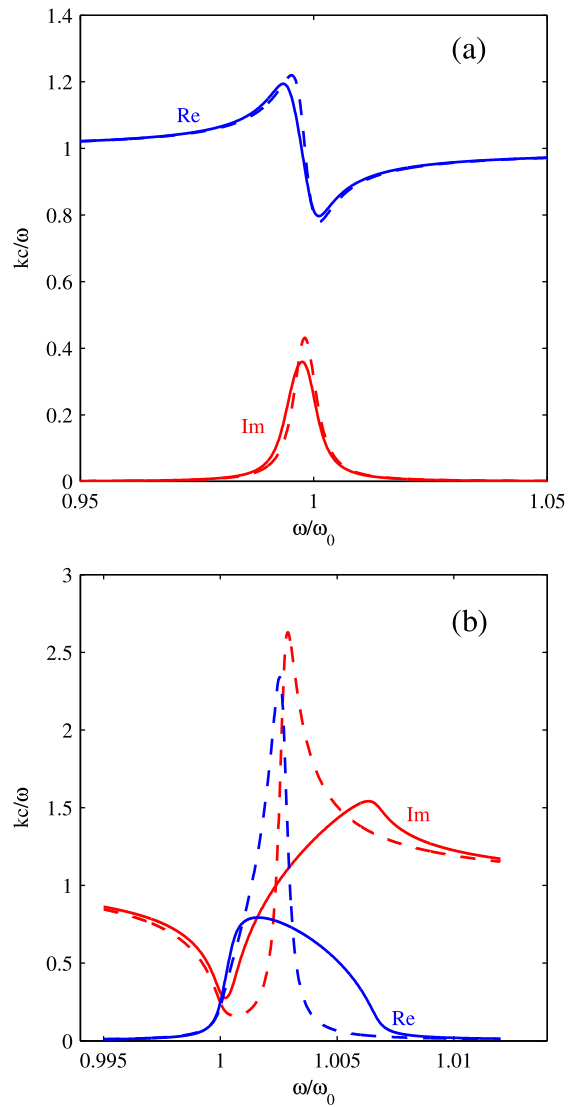}
\caption{\label{fig:fig5} Influence of the Lorentz-Lorenz local field correction on the dispersion properties of a wave propagating in a bulk toroidal metamaterial at $\bm{k}\bot\bm{n}$ and $\bm{E}||\bm{n}$. The full curves are obtained  with $\bm{H'}=\bm{H}+\bm{M}/3$ and the dashed curves - assuming $\bm{H'}=\bm{H}$ (low-density approximation). Real "Re" and imaginary "Im" parts of the normalized wavenumber $kc/\omega$. (a) Toroidal metamaterial with $\gamma/\omega_{0}=5.5\times10^{-3}$. (b) Composite toroidal metamaterial with $\gamma/\omega_{0}=5.5\times10^{-4}$ and $\omega_{p}/\omega_{0}=1.02\sqrt{2}$. In both (a) and (b) $\omega_{0}/\omega_{T}=0.07$ and $\omega_{0}=2$GHz. }
\end{figure}
The host media, in which the toroids are embedded is modeled with the effective permittivity tensor 
\begin{eqnarray}
    \widehat{\varepsilon}=\varepsilon\left(\hat{\delta}+(\varepsilon_{p}-1)\bm{n}\bm{n}\right),
    \label{eq:eleven}
\end{eqnarray}  
where $\hat{\delta}$ is the unit matrix and $\varepsilon_{p}=1$ pertains to the case of the host material being an isotropic dielectric. The combination $\varepsilon=1$ and $\varepsilon_{p}=1-\frac{k_p^{2}}{(\omega/c)^2-k_{z}^{2}}$, where $k_{p}$ is the effective "plasma" wavenumber \cite{17}, corresponds to a wire medium used as a host material. Note that at $k_{z}=0$ (wave propagation perpendicular to the wires) $\varepsilon_{p}$ reduces to $\varepsilon_{p}=1-\frac{\omega_p^{2}}{\omega^2}$. 

For an electromagnetic wave propagating in the bulk sample of the toroidal metamaterial sketched in Fig. \ref{fig:fig2} the electromagnetic field is that of an "extraordinary" wave, i.e. $\bm{E}=(E_{x}, 0, E_{z})$, $\bm{H}=(0, H_{y}, 0)$ and $\bm{k}=(k_{x}, 0, k_{z})$. The dispersion equation in this case is
\begin{eqnarray}
    \frac{\omega^{2}}{c^{2}}\varepsilon\varepsilon_{p}-k_{x}^2(1+\beta)-\varepsilon_{p}k_{z}^2=0,
   \label{eq:twelve}
\end{eqnarray} 
where
\begin{eqnarray}
    \beta=\frac{\omega^{4}\varepsilon\varepsilon_{p} }{\omega^{2}_{T}\left[\omega^{2}\left(1+\frac{k_{x}^{2}c^{2}}{3\omega_{T}^{2}}\right)-\omega^{2}_{0}+i\gamma \omega\right]}.
    \label{eq:thirteen}
\end{eqnarray}
 If the frequency $\omega$ is sufficiently far from the resonant frequency $\omega_{0}$ the contribution of the toroidal component can be neglected (i.e. $\beta=0$) and (\ref{eq:twelve}) becomes the standard dispersion equation of an extraordinary wave propagating in an uniaxial dielectric medium \cite{17}.  Note that the toroidal metamaterial does not interact with electromagnetic field, polarized perpendicular to $\bm{n}$.
 
Figure \ref{fig:fig5} illustrates the extent to which the effect of the mutual coupling between the toroidal metaparticles affects the dispersion properties of the wave propagating in a bulk sample of the metamaterial. As can be seen if the Q-factor $\omega_{0}/\gamma$ of an individual toroidal resonator is of the order or below 200, the mutual interaction between the metaparticles plays no role (Fig. \ref{fig:fig5}(a)), although the toroidal medium in the latter case can be formally regarded as "dense" (i.e. adjacent metapaticles have been brought in contact with each other and $\omega_{T}=\omega_{T0}$). This is because, by its nature, toroidal interaction with external electromagnetic fields is much weaker than electric- and magnetic-dipole interaction with the same field. The effective electromotive forces driving the currents in the toroidal solenoinds are, therefore, relatively small and the presence of stronger losses further limits the magnitude of these currents. Thus, the mutual interaction between toroidal moments is a higher-order effect even for "dense" toroidal media. In contrast, at higher Q-factor values (Fig. \ref{fig:fig5}(b)), the interaction between the metaparticles becomes significant.        
\begin{figure}
\includegraphics{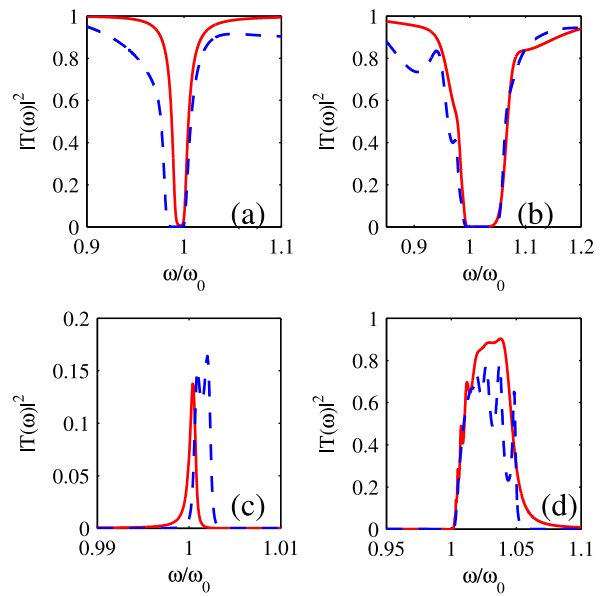}
\caption{\label{fig:fig6} Frequency dependence of the power transmission coefficient $|T(\omega)|^{2}$ for a metamaterial slab of thickness $l_{0}=10$cm at normal incidence and $\bm{E}||\bm{n}$. (a) Array of toroids with $d=5$mm, $R=d/3$, $\omega_{T}/2\pi=28.3$GHz, $\omega_{0}/2\pi=2.5$GHz and $\gamma/\omega_{0}=4.4\times10^{-3}$.  (b) Array of split-ring resonators with an effective permeability $\mu(\omega)=1+\frac{F\omega^{2}}{\omega_{0}^2-\omega^{2}-i\gamma\omega}$, where $F=0.1$. The other parameter values are the same as in (a). In both (a) and (b) the full (red) line corresponds to $\varepsilon=1$ and dashed (blue) line to $\varepsilon=2$. (c) Composite toroidal metamaterial (see Fig. \ref{fig:fig2}). The wires are assumed lossless and the other parameter values are the same as in (a). (d) Composite metamaterial consisting of an array of lossless wires and an array of split-ring resonators with the same effective permeability as in (b). In (c) and (d) the full (red) curve pertains to the case $\omega_{p}=\sqrt{2}\omega_{0}$, dashed (blue) curve - to $\omega_{p}=1.5\sqrt{2}\omega_{0}$ and $\gamma/\omega_{0}=4.4\times10^{-4}$.} 
\end{figure}
Figure \ref{fig:fig6} compares the electromagnetic properties of a toroidal metamaterial (a), (c) to that of an array of split-ring resonators (b), (d). As can be seen, toroidal response shows a strong dependence on the background dielectric permittivity $\varepsilon$ of the host material. In particular, doubling $\varepsilon$ effectively doubles the stop-band bandwidth of the toroidal medium (a), whereas doubling $\varepsilon$ has little effect on the response bandwidth of an array of split-ring resonators (b). This feature suggests possible sensor applications, since $\varepsilon$ depends on parameters such as temperature, stress, external electric field, etc. Thus small relative changes of $\varepsilon$ may result in significant relative changes of the transmission coefficient. The origin of this feature can be traced back to the \textit{magnetoelectric} nature of the toroidal response. Since $\bm{k \times H}=-\omega\varepsilon\varepsilon_{0}\bm{E}$, Eq. (\ref{eq:nine}) suggests that the macroscopic magnetization $\bm{M}$ is effectively driven by the electric field $\bm{E}$. Neglecting the interaction between the metaparticles, setting $k_{z}$ to zero and $\varepsilon_{p}$ to 1 in (\ref{eq:twelve}) results in $k^{2}=\frac{\omega^{2}}{c^{2}}\frac{\varepsilon}{1+g_{T}}$ with $g_{T}\approx\frac{\varepsilon \omega^{4}}{\omega_{T}^{2}(\omega^{2}-\omega_{0}^{2}+i\gamma\omega)}$. The quantity $\left|\frac{\varepsilon}{k}\frac{dk}{d\varepsilon}\right|$ (the ratio between the relative variation of the wavenumber  $k$ and the corresponding relative variation of the ambient permittivity $\varepsilon$) can be used to estimate the "sensitivity" of the system. The result is 
\begin{eqnarray}
    2\left|\frac{\varepsilon}{k}\frac{dk}{d\varepsilon}\right| = (1+2Re(g_{T})+\left|g_{T}\right|^{2})^{-\frac{1}{2}}.\nonumber
\end{eqnarray}
The resonant structure of $g_{T}(\omega)$ suggests the possibility that near the resonance $1+2Re(g_{T})+\left|g_{T}\right|^{2}\approx0$ and this results in a strong dependence of the power transmision coefficient $|T(\omega)|^{2}$ on $\varepsilon$. It should be emphasized that the latter result is a feature that is unique for toroidal media and does not exist for ordinary dielectric/magnetic materials composed of arrays of electric/magnetic dipoles. Indeed, since $k^{2}=\frac{\omega^{2}}{c^{2}}\varepsilon\mu$ in an ordinary dielectric/magnetic material and, hence, $g_{T}=0$, the "sensitivity" $\left|\frac{\varepsilon}{k}\frac{dk}{d\varepsilon}\right|$ remains relatively low at all frequencies in full accordance with Figure \ref{fig:fig6}(b) and (d).   
The Q-factor $\omega_{0}/\gamma$ of a single toroidal metaparticle in Fig. (\ref{fig:fig6})(a) is 230, which is comparable to the value of 50 measured by the authors of \cite{13} in the same frequency range, but for toroidal solenoids that are 10 times smaller than those considered here. At the same time the analysis indicates that higher Q-values should be achievable with larger solenoids \cite{13}. As Fig. (\ref{fig:fig6})(c) shows the same strong dependence of the toroidal medium response \textit{bandwidth} exists for a host medium with negative permittivity whereas no such feature is observed for arrays of split-ring resonators (d). In this case the stop bands from (a) and (b) are transformed into pass-bands by the presence of the negative permittivity material. Note, however, that much higher Q-factors are needed to produce a significant transmission in the composite toroidal medium.

Backward waves exist at frequencies within the pass-band shown in Fig. (\ref{fig:fig6})(c). To appreciate this consider the Poynting vector $\bm{S}$, in the absence of losses for $\bm{k}\bot\bm{n}$ and $\bm{E||n}$
\begin{eqnarray}
    \bm{S}=\frac{\bm{k|E|}^{2}}{2\omega \mu_{0}}\varepsilon(1-(\omega_{p}/\omega)^{2})(1+g(\omega)),
    \label{eq:fourteen}
\end{eqnarray}
\begin{figure}
\includegraphics{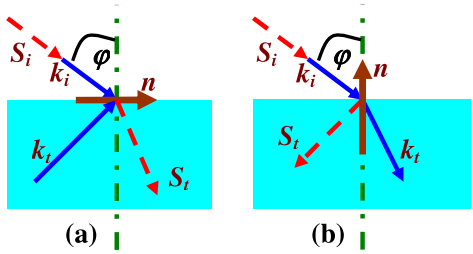}
\caption{\label{fig:fig7}Refraction properties of a composite toroidal metamaterial for a p-polarized incident wave.  Poynting vector $\bm{S}$ and wavevector $\bm{k}$ of the incident  ("\textit{i}") and the transmitted ("\textit{t}") wave. (a) The optic axis $\bm{n}$ is parallel to the interface.  (b) The optic axis is perpendicular to the interface.}
\end{figure}
where $g=\frac{\omega_{p}^{2}}{\omega^{2}-\omega_{p}^{2}}+\frac{\varepsilon \omega^{4}}{\omega_{T}^{2}(\omega^{2}(1+k^2c^2/(3\omega_{T}^{2}))-\omega_{0}^{2})}$ and $k^{2}=\frac{\omega_{2}}{c^{2}}\frac{\varepsilon}{1+g}$. This shows that if $1+g(\omega)>0$ and $\omega<\omega_{p}$ the material is transparent and the wavevector and the Poynting vector are antiparallel. 
Analysis similar to that of \cite{18} has been performed and the results are summarized in Fig. \ref{fig:fig7}. The components of the Poynting vector and the wavevector that are perpendicular to the optic axis point in opposite directions - i.e. $\bm{S_{\bot}.k_{\bot}}<0$. Depending on the orientation of the optic axis with respect to the interface this results in either positive or negative refraction, shown on Fig. \ref{fig:fig7}(a) and (b), respectively. Note, however, that despite the possibility of negative refraction (Fig. \ref{fig:fig7}(b)) the composite toroidal metamaterial is not a left-handed metamaterial \cite{19} since a wave propagating in a left-handed metamaterial is always a backward wave - it satisfies the more restrictive condition $\bm{S.k}<0$.    
 
In summary, it is shown that in a sharp contrast to materials that exist in nature, a new type of toroidal metamaterial shows a \textit{significant} toroidal response. The magnetoelectric nature of the toroidal response results in strong dependence of the transmission properties on the dielectric permittivity of the host medium, which suggests possible applications. Negative refraction and backward waves are found in a composite toroidal metamaterial. 

This work is supported by the
Engineering and Science Research Council (UK) under the Adventure
Fund and NanoPhotonics Portfolio Programmes.

\end{document}